# Modification of Surface Properties in Ruthenium Thin Films through Nitrogen Ion Irradiation

Anmol Sharma[a], Arnab Tripathy[b], Rajeev Gupta[b], Raj Kumar[c], Harsh Vardhan[d], Ratnesh Kumar Pandey[e], Atul Thakur[a], Shalendra Kumar[b], Ranjeet Kumar Brajpuriya[b], Vishakha Kaushik[f], and Sachin Pathak[b]*

[a]Centre of Excellence in Nanotechnology, Lovely Professional University, Phagwara, Punjab 144411, India
[b]Department of Physics, School of Advanced Engineering, UPES Dehradun, Uttarakhand, 248007, India
[c]Inter University Accelerator Centre, Aruna Asif Ali Marg, New Delhi 110067, India
[d]Amity Centre for Spintronic Materials, Amity University Noida, India, 201313
[e]Applied Sciences and Humanities Department, Samrat Ashok Rajkiya Engineering College, Mirzapur, 231001, Uttar Pradesh, India
[f]Materials & Nano Engineering Research Laboratory, Department of Physics, School of Physical Sciences, DIT University, Dehradun- 248009, India

*Corresponding author: s.pathak@ddn.upes.ac.in (Sachin Pathak)

**Abstract:** Surface modification is an important strategy for tailoring the physical and chemical properties of materials for advanced technological applications. In thin films, changes in surface structure, roughness, and composition can strongly influence their functional behavior. In this work, we studied the effect of $N^+$-ion irradiation at 20 keV energy on Ru thin films of two different thicknesses, 100 Å and 1000 Å, over a fluence range of ~$10^{14}$ to $10^{17}$ ions/cm$^2$. The main objective of this study is to investigate the evolution of the structural and wettability properties of Ru thin films of different thicknesses under increasing ion fluence. Overall, the results reveal that increasing ion fluence promotes surface roughening in both film thicknesses, which is correlated with an increase in contact angle. This indicates a clear transition from hydrophilic to hydrophobic behaviour, with the highest contact angle of ~95.54° observed at a fluence of $10^{17}$ ions/cm$^2$. TRIM simulations reveal that for the 100 Å film, the projected ion range exceeds the film thickness, allowing interaction with the bottom substrate, whereas in the 1000 Å film, ion penetration remains largely confined within the Ru layer. GIXRD confirms fluence-dependent structural modification, including peak broadening with preferred orientation changes from Ru (101) to (002). Overall, this study demonstrates that $N^+$-ion irradiation serves as an effective route for tuning both the surface morphology and wettability of Ru thin films, enabling controlled enhancement of hydrophobicity.

## Highlights

- Nitrogen ion irradiation directly tailors the surface morphology of Ru thin films.
- Increased ion fluence causes significant nanoscale surface roughening and structural evolution.
- N-ion irradiation-induced nanostructuring successfully modifies a hydrophilic-to-hydrophobic transition.

## Graphical abstract

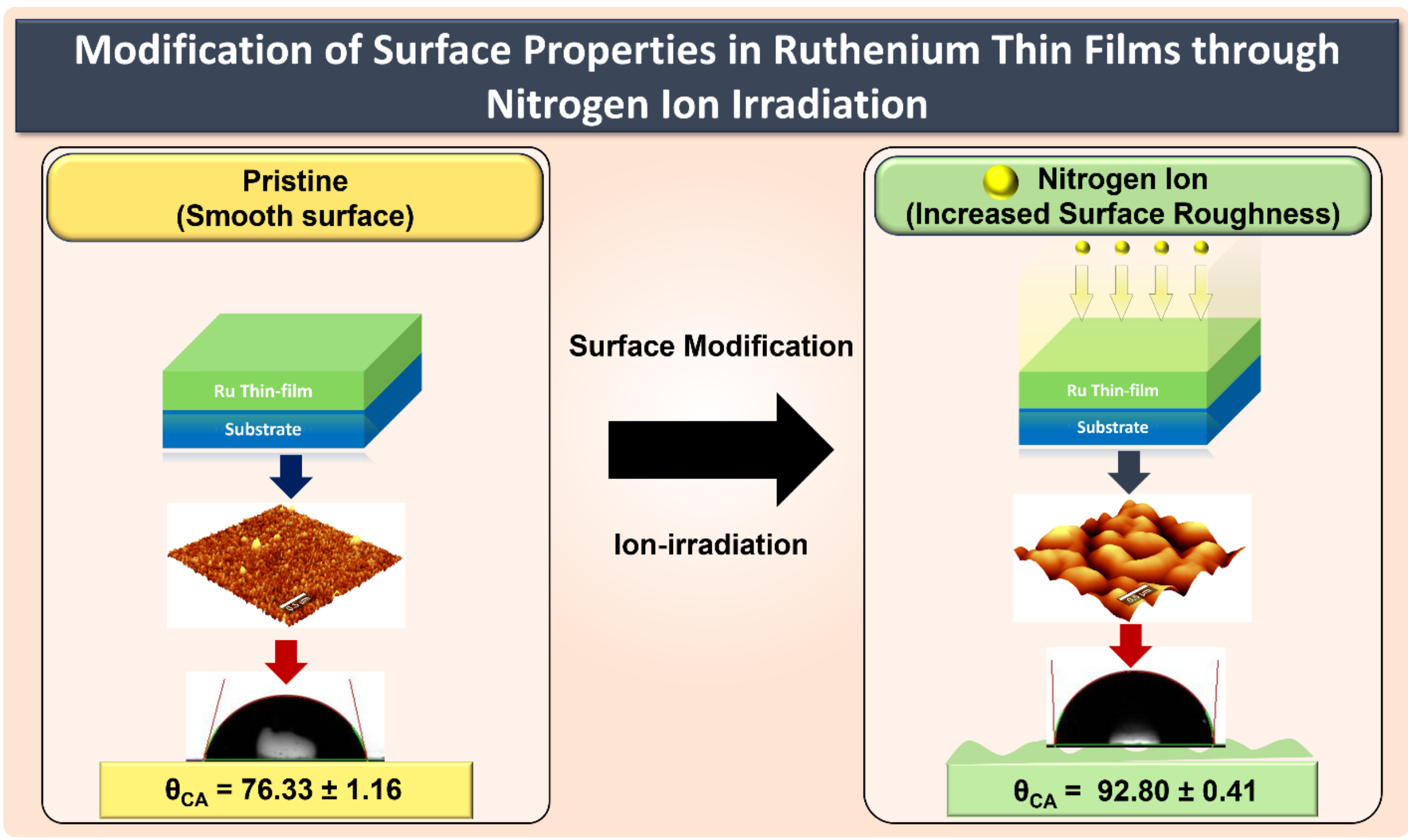

## 1. Introduction

Functional noble metal has emerged as an important field of research because of their exceptional properties like chemical stability, and tunable surface properties for a wide range of applications, including microelectronics, protective coatings, and energy devices[1–3]. In many of these applications, the functional performance of these materials is strongly influenced by its roughness and surface morphology. Therefore, surface nanostructuring provides an effective approach for tailoring the surface properties of thin films through controlled modification of their morphology. However, achieving nanoscale or microscale surface features directly on thin films can be challenging using conventional fabrication methods. Although various approaches, including lithography, metal-assisted etching, and plasma-based processes, have been studied for generating nanostructured surfaces, these techniques often involve multiple processing steps, additional masking or chemical treatments[4–6].

In comparison, ion-irradiation offers a relatively direct and mask-free route for nanostructuring and tailoring surface properties[7,8]. Through processes like defect creation, sputtering, and nanoscale transformation, ion irradiation allows for precise control of surface properties, resulting in significant modifications to surface chemistry and topography. These changes directly influence the surface energy of the material, thereby modifying its wetting behaviour. It is well established that ion-irradiation can alter the surface properties of metallic and semiconductor thin films through nano-structuring and compositional changes, leading to hydrophilic-to-hydrophobic transitions[9,10]. These findings highlight the relationship between surface morphology, chemical composition, and wettability, establishing ion-irradiation as a powerful method for tailoring surface properties at the nanoscale.

Among heavy transition metals, Tantalum (Ta), Platinum (Pt) and Ruthenium (Ru), has attracted significant attention because of their high utilization in electronic applications, but are often susceptible to environmental degradation due to moisture adsorption and surface contamination[11–14]. To address these challenges, considerable efforts have been made toward engineering protective surface layers through alloying or compound formation[15,16]. Transition metal nitrides, in particular, are known for their enhanced chemical stability, mechanical robustness, and hydrophobic characteristics[17–20]. Nitrogen species offers a unique advantage in this regard, as it enables simultaneous physical and chemical modification of the surface[21–24]. In addition to inducing structural rearrangements, nitrogen incorporation can modify the chemical environment and promote the formation of nitride phases, thereby modifying the surface energy and chemical reactivity. This combined effect of morphological evolution and chemical modification makes incorporation of nitrogen in other transition metals a promising strategy for tailoring surface properties of thin-film systems.

However, systematic studies on ion irradiation-induced wettability modification in transition metal thin films deposited via sputtering remain relatively limited. Ru, owing to its excellent chemical stability, corrosion resistance, and compatibility with thin-film technologies, is a material of considerable technological importance. Previous studies have shown that Ru-based coatings exhibit strong dependence of surface properties on processing conditions[25]. For example, Ru ion implantation on stainless steel highlights that parameters such as implantation dose, energy, and surface roughness significantly influence corrosion resistance[26]. However, Ru surface response to nitrogen ion irradiation, particularly in terms of wettability modulation, has not been comprehensively investigated. Therefore, in this work, we aim to explore the effect of nitrogen ion-irradiation at different fluences on the surface properties of Ru thin films, with a focus on understanding the underlying mechanisms governing irradiation-induced changes in surface properties.

The sections of this paper are organized as follows. Section 2 describes the experimental methodology, including details of sample preparation, $N^+$-ion irradiation, and the characterization techniques used. Section 3 presents and discusses the results, focusing on the impact of ion irradiation on the structural properties and wettability of the films, particularly the observed changes in contact angle. Finally, Section 4 summarizes the main findings and outlines possible directions for future research on ion irradiation-induced surface modifications for related applications.

## 2. Experimental details

Thin films of Ru with two different thicknesses, 100 Å and 1000 Å, were deposited on $Si/SiO_2$ (100) substrates. The films were grown at room temperature using a DC magnetron sputtering system operating under ultra-high vacuum conditions of ~$4.2 \times 10^{-7}$ Torr. To eliminate any residual surface contamination, the Ru target was pre-sputtered for 10 minutes before deposition. The sputtering process was carried out in a controlled argon (Ar) environment with a constant flow rate of 10 sccm, while maintaining a sputtering power of 100 W. Substrate rotation at 15 rpm was employed throughout the deposition process to ensure homogeneous film thickness and surface uniformity across the substrate. After deposition, the Ru films were irradiated with nitrogen ions ($N^+$) with an energy of 20 keV at normal incidence. The irradiation experiments were performed using a tabletop accelerator at the Inter-University Accelerator Centre (IUAC), New Delhi. Four different fluences of ~$10^{14}$, $10^{15}$, $10^{16}$, and $10^{17}$ ions/cm$^2$ were used to irradiate the Ru films. During irradiation, the chamber pressure was maintained at better than ~$9.5 \times 10^{-7}$ Torr to minimize contamination effects.

The thickness and surface roughness of both as-deposited and ion-irradiated Ru thin films were examined using X-ray reflectivity (XRR). Phase evolution and structural modifications induced by ion irradiation were further studied using grazing incidence X-ray diffraction (GIXRD). To ensure reliable XRR data, a knife-edge was employed to minimize the footprint effect, and measurements were collected with an angular step size of 0.01°. The measured reflectivity data were analyzed using GenX software[27] (version 3.6.16), which enabled determination of film thickness and surface roughness. GIXRD measurements were carried out over the 2θ range of 30-90° with a step size of 0.05°. The incident angle was fixed at 0.5° to reduce the substrate contributions. Both XRR and GIXRD experiments were performed on a Panalytical Empyrean diffractometer (Model: Emp3, Panalytical, Netherlands), equipped with Cu-$K_\alpha$ radiation ($\lambda$ = 1.5405 Å) generated at 45 kV and 40 mA. Surface topography of as-deposited and irradiated thin films was evaluated by atomic force microscopy (AFM) using a Digital Instruments Nanoscope IIIa (AFM; Santa Barbara, CA, USA). Imaging was performed in tapping mode using a silicon probe. While the wetting dynamics of ruthenium surface features in pristine and $N^+$-ion irradiated states were investigated using water droplet contact angle measurements (KRUSS DSA-10). As the final characterization, Field emission scanning electron microscopy (FESEM) was also performed (JEOL JSM-7610F-PLUS). Prior to imaging, the samples were coated with gold for 2 minutes to mitigate surface charging. Thin-film deposition, $N^+$-ion irradiation, and all other measurements were performed at room temperature.

## 3. Results and discussion

In order to systematically examine how ion irradiation modifies surface properties, we have studied two different Ruthenium film thicknesses, 100 Å and 1000 Å. Figure 1(a) and (b) show the schematic representation of these films, which were irradiated with N-ions at 20 keV energy. To understand the ion-matter interaction, TRIM simulations[28] were performed using 50,000 incident ions to obtain statistically reliable results. For the 100 Å Ru film, the projected ion range was ~460 Å, indicating that N-ions penetrate through the Ru layer and extend into the Si substrate, as shown in ion distribution plot in Figure 1(c). In contrast, for the 1000 Å film, the ion range was ~230 Å (shown in Figure 1 (d)), confirming that the ions remain confined within the Ru layer.

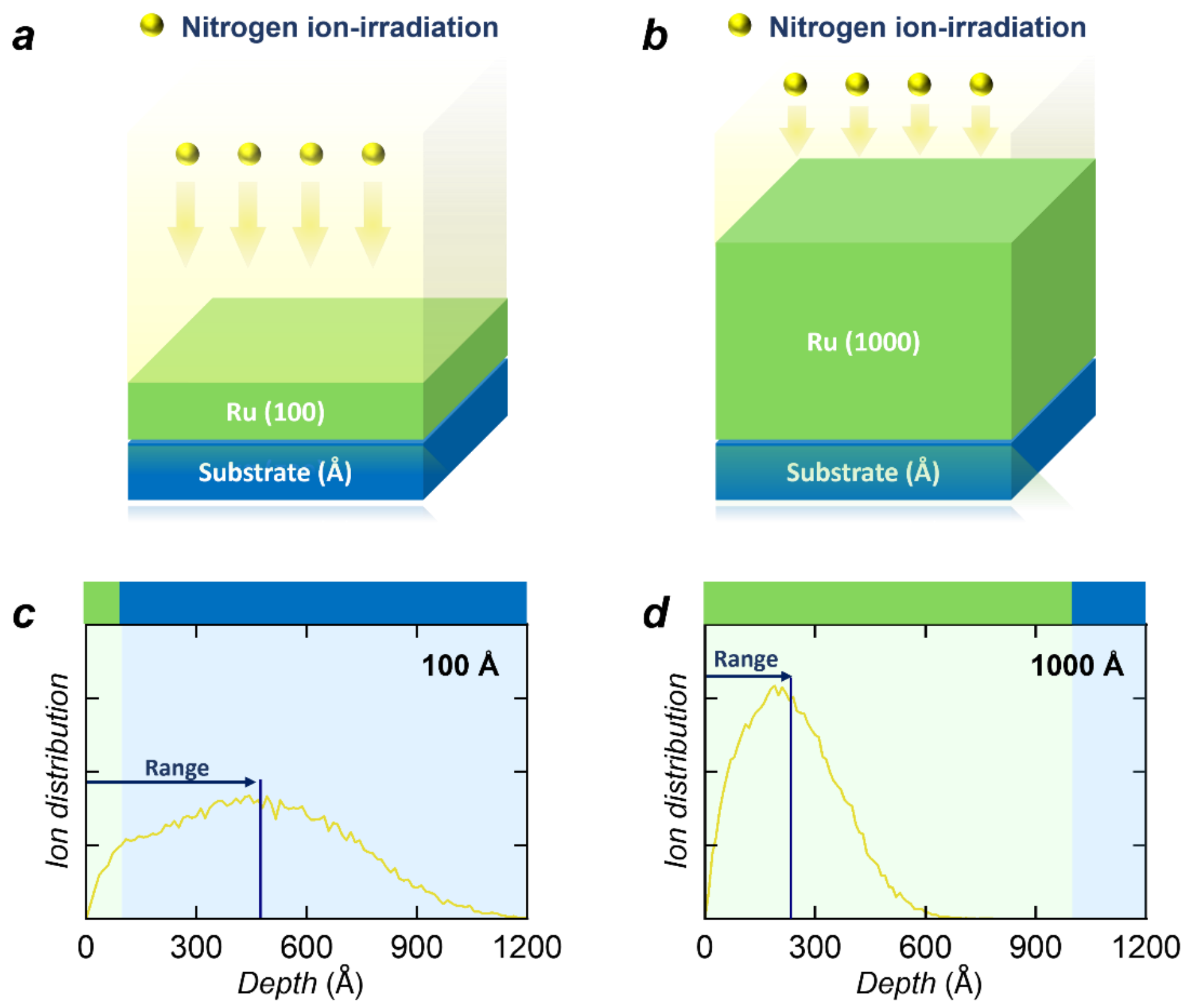

**Figure 1:** Schematic illustration of **(a)** 100 Å and **(b)** 1000 Å Ru films under 20 keV $N^+$-ion irradiation. **(c, d)** TRIM-simulated profiles showing $N^+$-ion penetration in 100 Å and 1000 Å Ru films, respectively, with the projected $N^+$-ion range indicated.

For the 100 Å Ru film, the calculated sputtering yield of Ru is approximately 1.19 atoms/ion, while the contribution from the substrate remains negligible (~0.00038 atoms/ion). Increasing the Ru thickness to 1000 Å results in slight increase in the Ru sputtering yield, reaching approximately 1.28 atoms/ion. The damage analysis shows a similar trend, with the 100 Å film exhibiting ~346 atomic displacements/ion and ~320 vacancies/ion, compared with ~337 displacements/ion and ~314 vacancies/ion for the 1000 Å film. Although the overall defect generation is comparable for both thicknesses, the distribution of the defects in both the film thicknesses differs considerably because of the different N-ion penetration depths. In the thinner Ru film, a fraction of the incident $N^+$-ions can penetrate through the Ru layer and reach the underlying substrate, promoting atomic redistribution and defect formation at the interface of both the materials. In contrast, the thicker Ru film effectively confines the $N^+$-ion distribution within the Ru layer, resulting in localized distribution of defects and structural disorder within the Ru film. As a result, although the overall ion-induced damage is similar for both film thicknesses, the damage in the thicker film is primarily limited to the Ru layer, while the thinner Ru film undergoes a larger atomic redistribution as a result of $N^+$-ion transmission into the substrate.

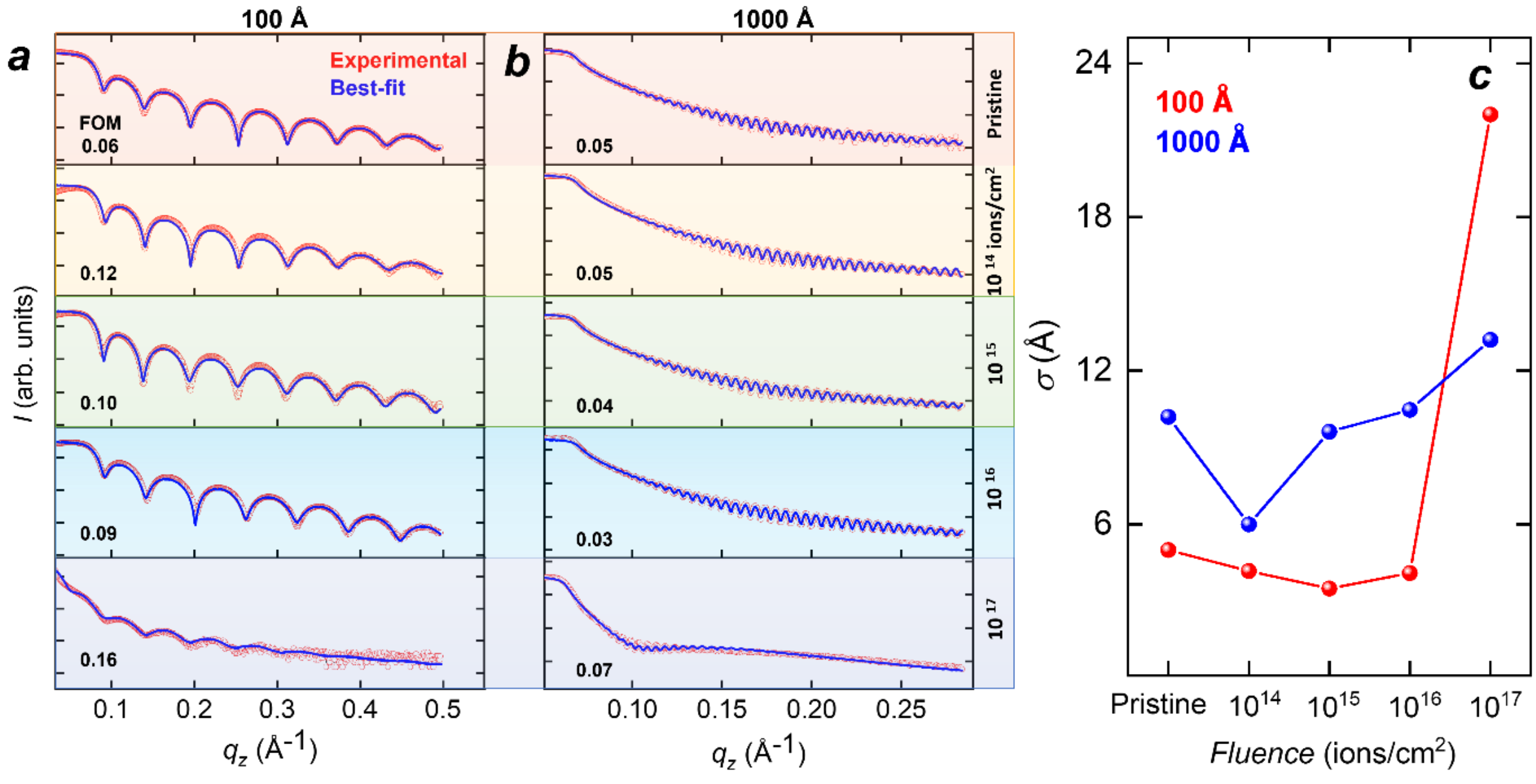

**Figure 2:** XRR data (red symbols) with corresponding fitting curves (blue lines) for Ru thin films of **(a)** 100 Å thickness and **(b)** 1000 Å thickness (bottom left of each plot shows figure of merit (FOM) values). Results are presented for pristine and $N^+$-irradiated thin films at fluences of $1\times10^{14}$, $1\times10^{15}$, $1\times10^{16}$ and $1\times10^{17}$ ions/cm$^2$ at 20 keV energy. **(c)** Surface roughness variation (σ in ±3 Å) for Ru films having thicknesses of 100 Å and 1000 Å as a function of $N^+$-ion fluence ranging from $1\times10^{14}$ to $10^{17}$ ions/cm$^2$ at 20 keV energy.

The thickness and surface roughness of both the as-deposited and $N^+$-ion-irradiated Ru thin films were determined using XRR measurements. Figure 2 (a) and (b) show the XRR patterns for both films in the pristine as well as after $N^+$-ion irradiation at different fluences ranging from $10^{14}$ to $10^{17}$ ions/cm$^2$. The red curves represent the experimental data, while the blue curves correspond to the fitted profiles. For as-deposited 100 Å Ru film, thickness calculated was ~103 Å, with an initial surface roughness of ~5 Å. After $N^+$-ion irradiation, the surface roughness showed a non-monotonic evolution with increasing $N^+$-ion fluence. It initially decreased, reaching a minimum at a fluence of $10^{15}$ ions/cm$^2$, and after which, it was increased progressively reaching a maximum value of ~22 Å at fluence of $10^{17}$ ions/cm$^2$, indicating ion-induced modification of the film surface. A similar trend was observed for the thicker 1000 Å Ru film (pristine ~1045 Å) , where the surface roughness initially decreased to ~6 Å and then increased to ~13 Å at a fluence of $10^{17}$ ions/cm$^2$. Figure 2 (c) shows the variation in surface roughness ($\sigma$) of both Ru films as a function of $N^+$-ion fluence ranging from $10^{14}$ to $10^{17}$ ions/cm$^2$.

Apart from surface roughness of these films, critical angles ($\theta_c$) also showed some variations at high fluences. In both films, $\theta_c$ remained close to that of bulk Ru ($q_c$~0.068 Å$^{-1}$) up to a fluence of $10^{16}$ ions/cm$^2$. However, at the highest fluence of $10^{17}$ ions/cm$^2$, a noticeable shift in the critical angle was observed towards lower angles of ~0.041 Å$^{-1}$ and 0.062 Å$^{-1}$ for 100 Å and 1000 Å Ru film respectively, indicating significant changes in both the film. Since the critical angle in XRR is closely related to the material's density ($\delta$) [$\theta_c \propto \sqrt{\delta}$], this shift suggests that the density of the Ru layers is reduced at higher irradiation fluence. Previous studies have also demonstrated similar results where ion irradiation can lead to structural transformations as well as density variations in thin films[29]. To further explore these possible structural or phase changes in the present case, GIXRD measurements were carried out.

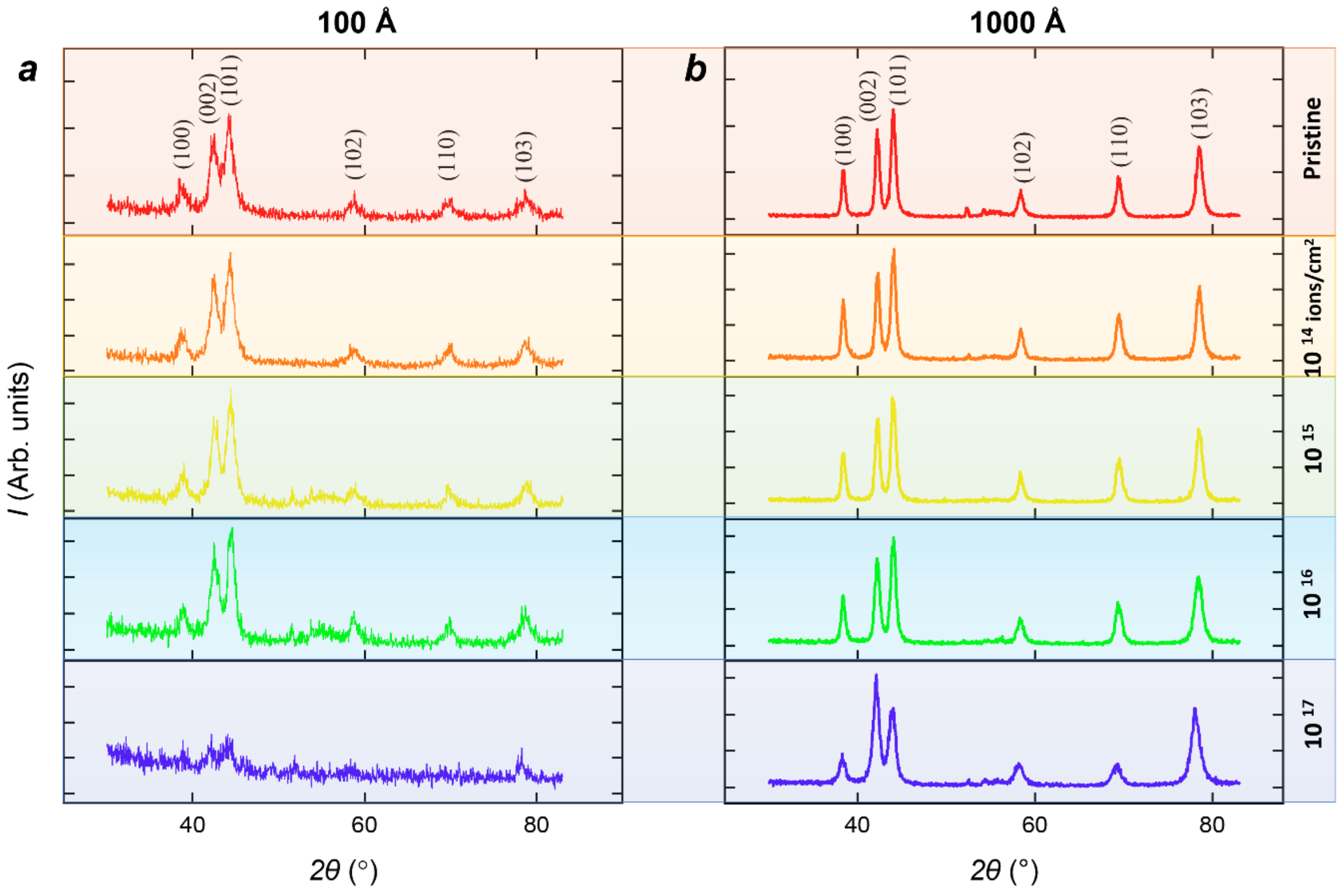

**Figure 3:** GIXRD patterns of Ru thin films with thicknesses of **(a)** 100 Å and **(b)** 1000 Å in the as-deposited state and after $N^+$-ion irradiation at different fluences at 20 keV energy.

Figure 3 shows the GIXRD patterns for both Ru films in the as-deposited and irradiated states, with the corresponding phases marked in the figure. For the as-deposited Ru films, several diffraction peaks are observed at around ~38°, ~42°, ~45°, ~58°, ~70°, and ~78°, which can be indexed to the (100), (002), (101), (102), (110), and (103) planes, respectively, confirming the polycrystalline nature of the films[30]. In both thicknesses, noticeable changes are observed in the relative intensities of the (002) and (101) peaks upon irradiation. In the pristine films, the (101) peak is dominant; however, with increasing $N^+$-ion fluence, its intensity decreases relative to the (002) peak. This behavior suggests a modification in the preferred orientation or texture of the Ru film induced by $N^+$-ion irradiation. Apart from these changes, the positions and intensities of the other diffraction peaks remain largely unaffected, indicating that the overall crystal structure remain same. Furthermore, no new diffraction peaks corresponding to secondary or nitrogen-related phases are observed, confirming the phase stability of the Ru films after $N^+$-ion irradiation. For the100 Å Ru film, distinct behavior was observed at the highest fluence of $10^{17}$ ions/cm$^2$, where the diffraction peaks were almost disappeared. This indicates significant ion-induced atomic disorder and possible amorphization of the film. This observation is consistent with the XRR results, where a reduction in the $\theta_c$ at this fluence was observed, suggesting a decrease in film density.

In addition to the changes in peak intensity, (002) and (101) peaks shows a small shift toward lower 2θ angles as the ion fluence increases. For the 100 Å film, for example, the (002) peak shifts from 42.41° to 42.07°, while the (101) peak moves from 44.26° to 44.08°. A similar shift is also observed for the 1000 Å film. This movement toward lower 2θ angles indicates an increase in the lattice d-spacing with relatively small lattice strain (under 1%), suggesting irradiation-induced lattice expansion with increasing ion fluence. Together, these results suggest that high-fluence $N^+$-ion irradiation leads to substantial structural modification, including disordering and density reduction, particularly in thinner Ru films. Therefore, to systematically quantify the microstructural evolution, detailed peak fitting of the dominant (002) and (101) peak reflections was performed calculate the variations in crystallite size.

Figure 4 (a) and (b) show the peak fitting of the (002) and (101) peaks obtained from the GIXRD patterns for the as-deposited Ru films for 100 and 1000 Å thicknesses, respectively. The peaks were fitted using a pseudo-Voigt function to accurately determine their full width at half maximum (FWHM). Similarly, all other relevant peaks were also fitted using the same approach to extract the FWHM values, which were then used to estimate the crystallite size (*D*) of the films using Scherrer's equation[31]:

$$D = \frac{k\lambda}{\beta cos\theta} \quad (1)$$

where λ is the X-ray wavelength, $\beta$ is the FWHM of the XRD peak, *D* is the crystallite size, and *k* is the Scherrer constant (*k*=0.9). Comparing the as-deposited states, the 1000 Å film exhibits a significantly larger initial crystallite size compared to the 100 Å film, indicating improved initial crystallinity in the thicker film. Upon irradiation, the 1000 Å Ru film shows a systematic decrease in crystallite size with increasing $N^+$-ion fluence. For the (002) and (101) orientations, D decreases gradually from 152 Å to 106 Å and from 140 Å to 95 Å, respectively. This progressive reduction in crystallite size reflects the steady accumulation of irradiation-induced structural disorder and partial amorphization.

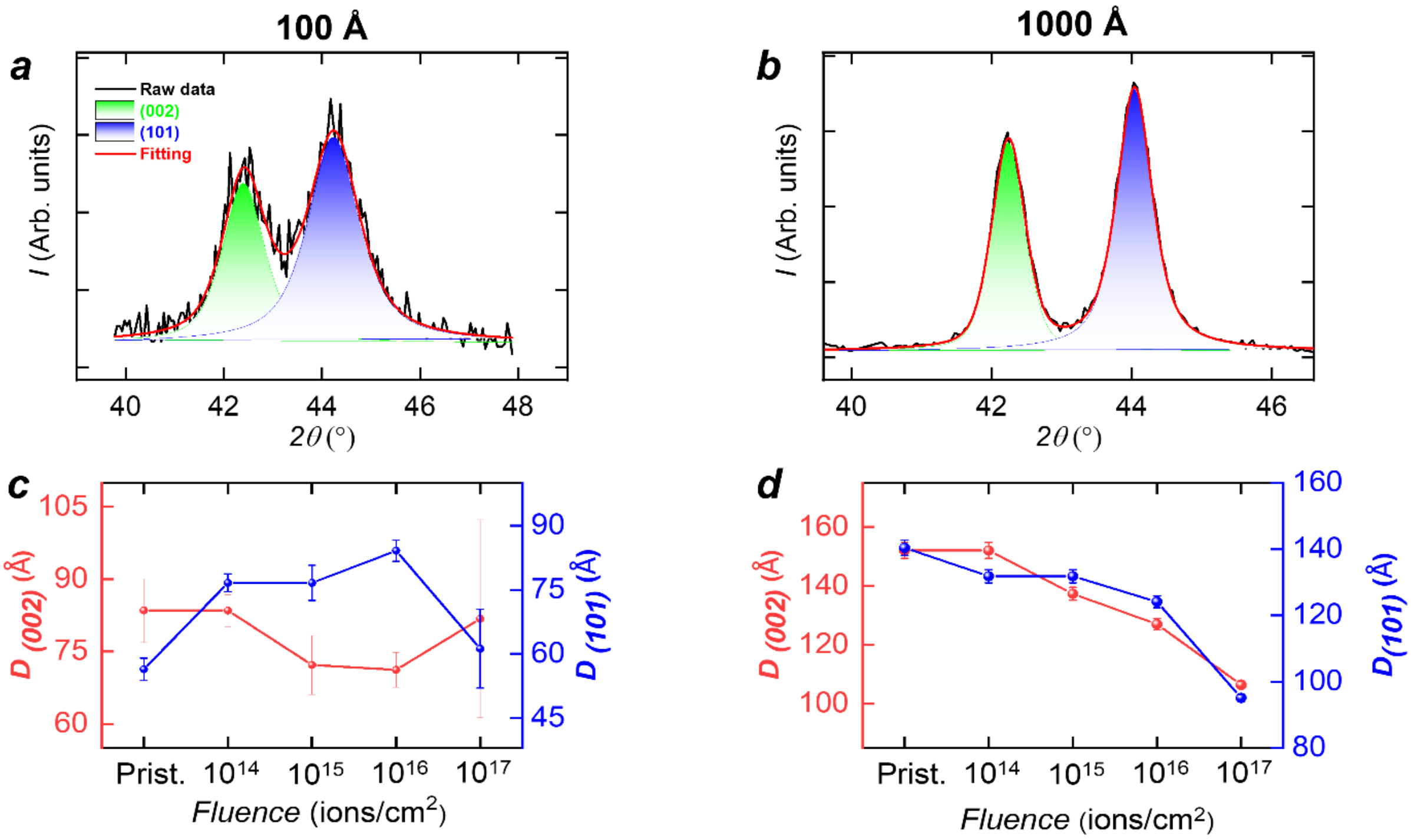

**Figure 4:** Peak fitting of the (002) and (101) peaks for **(a)** 100 Å and **(b)** 1000 Å Ru films, respectively, where black curves represent the experimental data and red curves show the fitted profiles. **(c), (d)** Variation in crystallite size (*D*) for the (002) and (101) orientations as a function of $N^+$-ion fluence at 20 keV energy for both films.

In contrast, the 100 Å Ru film exhibits a less systematic variation in crystallite size at intermediate fluences, followed by a substantial structural degradation at the highest fluence of $10^{17}$ ions/cm$^2$. The stronger structural modification observed in the thinner film can be attributed to the relatively greater penetration depth of the $N^+$-ions. In the 100 Å film, ions can more easily reach the underlying substrate, promoting atomic redistribution and defect generation across the film-substrate interface. These effects align perfectly with the XRR measurements for the 100 Å film, which revealed a more pronounced reduction in the $\theta_c$ and a larger increase in surface roughness at the highest fluence. The combined GIXRD and XRR results clearly demonstrate that the thinner Ru film undergoes substantially stronger irradiation-induced structural modification than the thicker film, particularly at high $N^+$-ion fluences. Table 1 summarizes the corresponding values of peak position, FWHM, and crystallite size, showing the changes in crystallite size with increasing N-ion fluence for both film thicknesses.

**Table 1:** Extracted peak positions, FWHM, and calculated crystallite size *D* (in Å) for the (002) and (101) planes of Ru films at different $N^+$-ion fluences.

**Ru - 100 Å**

| Sr. No. | Condition (ions/cm$^2$) | (002) | | | (101) | | |
|---|---|---|---|---|---|---|---|
| | | ***2θ* (º)** | ***FWHM* (º)** | ***D* (Å)** | ***2θ* (º)** | ***FWHM* (º)** | **~*D* (Å)** |
| 1. | Pristine | 42.41 ± 0.03 | 1.02 ± 0.08 | 83.5 ± 6.6 | 44.26 ± 0.02 | 1.52 ± 0.07 | 56.4 ± 2.6 |
| 2. | $1\times10^{14}$ | 42.48 ± 0.01 | 1.02 ± 0.04 | 83.5 ± 3.3 | 44.36 ±0.01 | 1.12 ± 0.03 | 76.6 ± 2.1 |
| 3. | $1\times10^{15}$ | 42.52 ± 0.03 | 1.18 ± 0.10 | 72.2 ± 6.1 | 44.41 ± 0.02 | 1.12 ± 0.06 | 76.6 ± 4.1 |
| 4. | $1\times10^{16}$ | 42.60 ± 0.02 | 1.19 ± 0.06 | 71.6 ± 3.6 | 44.49 ± 0.01 | 1.02 ± 0.03 | 84.1 ± 2.5 |
| 5. | $1\times10^{17}$ | 42.07 ± 0.08 | 1.04 ± 0.26 | 81.8 ± 20.5 | 44.08 ± 0.07 | 1.40 ± 0.21 | 61.2 ± 9.2 |

**Ru - 1000 Å**

| Sr. No. | Condition | (002) | | | (101) | | |
|---|---|---|---|---|---|---|---|
| | | ***2θ* (º)** | ***FWHM* (º)** | **~*D* (Å)** | ***2θ* (º)** | ***FWHM* (º)** | **~*D* (Å)** |
| **1.** | Pristine | 42.22 ± 0.01 | 0.56 ± 0.01 | 152.0 ± 2.7 | 44.02 ± 0.01 | 0.61 ± 0.01 | 140.4 ± 2.3 |
| **2.** | $1\times10^{14}$ | 42.25 ± 0.01 | 0.56 ± 0.01 | 152.0 ± 2.7 | 44.04 ± 0.01 | 0.65 ± 0.01 | 131.8 ± 2.0 |
| **3.** | $1\times10^{15}$ | 42.24 ± 0.01 | 0.62 ± 0.01 | 137.3 ± 2.2 | 44.05 ± 0.01 | 0.65 ± 0.01 | 131.8 ± 2.0 |
| **4.** | $1\times10^{16}$ | 42.22 ± 0.01 | 0.67 ± 0.01 | 127.0 ± 1.9 | 44.02 ± 0.01 | 0.69 ± 0.01 | 124.1 ± 1.8 |
| **5.** | $1\times10^{17}$ | 42.10 ± 0.01 | 0.80 ± 0.01 | 106.4 ± 1.3 | 43.91 ± 0.01 | 0.90 ± 0.01 | 95.1 ± 1.1 |

Figure 5 (a)-(d) present the surface morphology of Ru films in pristine and after $N^+$-ion irradiation at different fluences. As seen in Figure 5(a), the surface of the 100 Å Ru film undergoes significant modification with increasing ion fluence. The surface roughness increased from ~6 Å in the pristine state to ~40 Å at the highest fluence of $10^{17}$ ions/cm$^2$. This increase in roughness can be attributed to enhanced sputtering effects at higher fluences, particularly due to the low film thickness, with a sputtering yield of ~1.19 atoms/ion for 20 keV $N^+$-ions as estimated from TRIM simulations. These observations are consistent with the XRR and GIXRD results, where notable changes such as variation in critical angle (indicating density modification) and reduced peak intensity were observed at higher fluences. The FESEM images in Figure 5(c) also shows that for the 100 Å film, surface goes from a uniform to a broken, disconnected structure with island-like formations at the highest dose.

A similar trend in roughness is observed for the thicker 1000 Å Ru film (shown in Figure 5 (b)), where the surface roughness increases from ~10 Å to ~13 Å with increasing fluence. However, the extent of modification is comparatively smaller which can be due to the higher film thickness. The FESEM images in Figure 5(d) clearly shows that the 1000 Å film avoids the severe structural changes seen in the thinner film, even at the maximum fluence. Hence, these irradiation-induced changes in surface morphology are expected to directly influence the surface properties of the films, which is further analyzed using drop shape (contact angle $\theta_{CA}$) measurements.

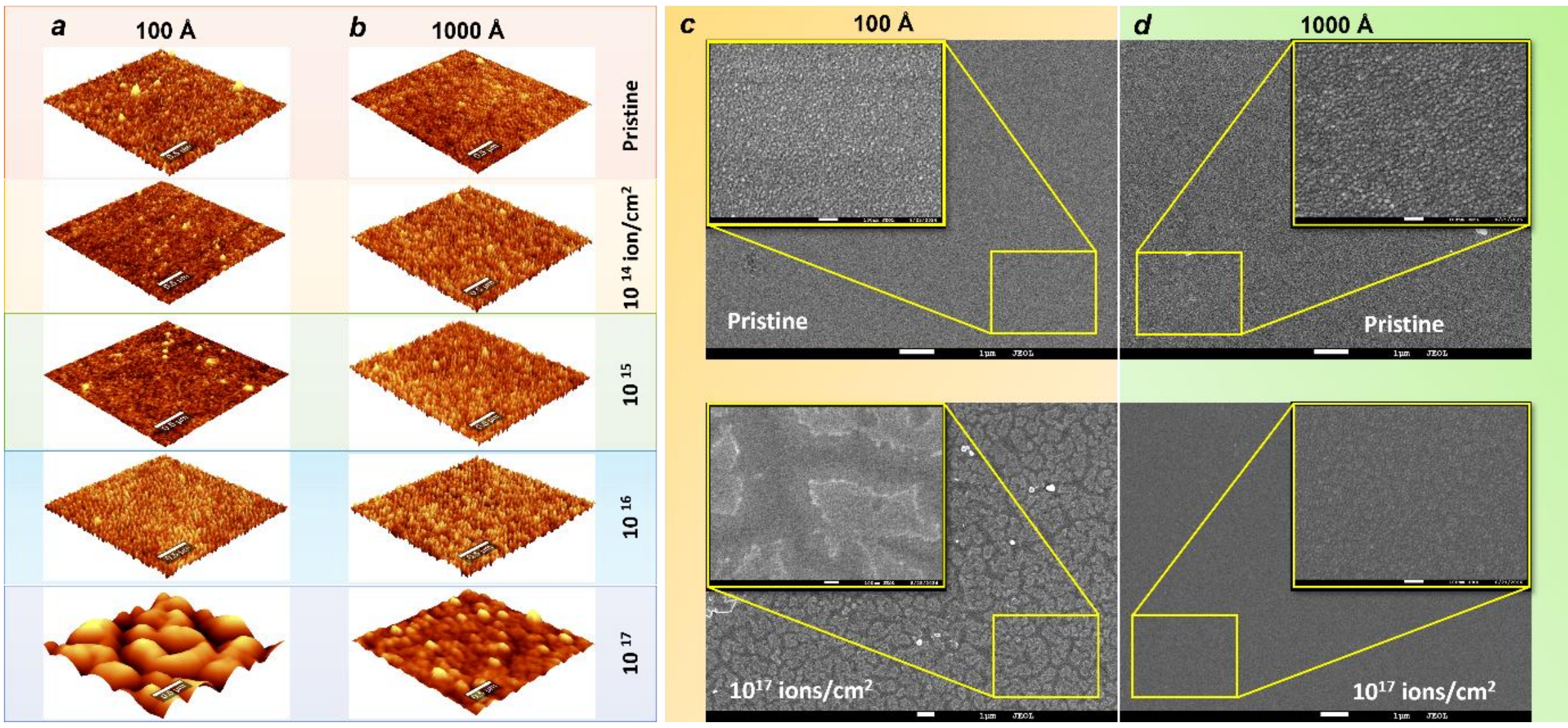

**Figure 5:** Surface morphology of Ru thin films. AFM images of the **(a)** 100 Å and **(b)** 1000 Å films in the pristine and after N-ion irradiation at various fluences. FESEM images comparing the pristine and the highest irradiation fluence of $10^{17}$ ions/cm$^2$ for the **(c)** 100 Å and **(d)** 1000 Å films (scale bars: 1 $\mu$m, insets: 100 nm).

Figure 6 presents the contact angle measurements on Ru films before and after N-ion irradiation. To ensure uniformity, a set volume of 11 $\mu L$ was used for all measurements in the drop shape analysis utilizing DI water droplets. This volume corresponds to the point at which the droplet detaches from the needle under gravity, providing a reproducible condition for comparison. The results show a clear dependence of wettability on $N^+$-ion irradiation. For the 100 Å Ru film, in the pristine state exhibits a $\theta_{CA}$ of 87.20±0.35° which was greater than that of the 1000 Å Ru film of 76.33±1.16°. For both film thicknesses, the $\theta_{CA}$ increases systematically with increasing N-ion fluence, indicating a gradual transition toward more hydrophobic behaviour. For the 100 Å Ru film, $\theta_{CA}$ increased from 87.20±0.35° to 95.54±2.23° at the highest fluence of $10^{17}$ ions/cm$^2$, while for the 1000 Å film, it increased from 76.33±1.16° to 92.80±0.41°. Figure 6 (c) shows the variation in contact angle ($\theta_{CA}$) for both Ru films having thicknesses of 100 Å and 1000 Å as a function of $N^+$-ion fluence, with a transition from hydrophilic to hydrophobic nature[32]. This observed behaviour can be directly correlated with the irradiation-induced changes in surface morphology, where increased roughness and irradiation induced redistribution of atoms at the surface leads to possible amorphization of the Ru film leading to change in the surface energy and wetting characteristics.

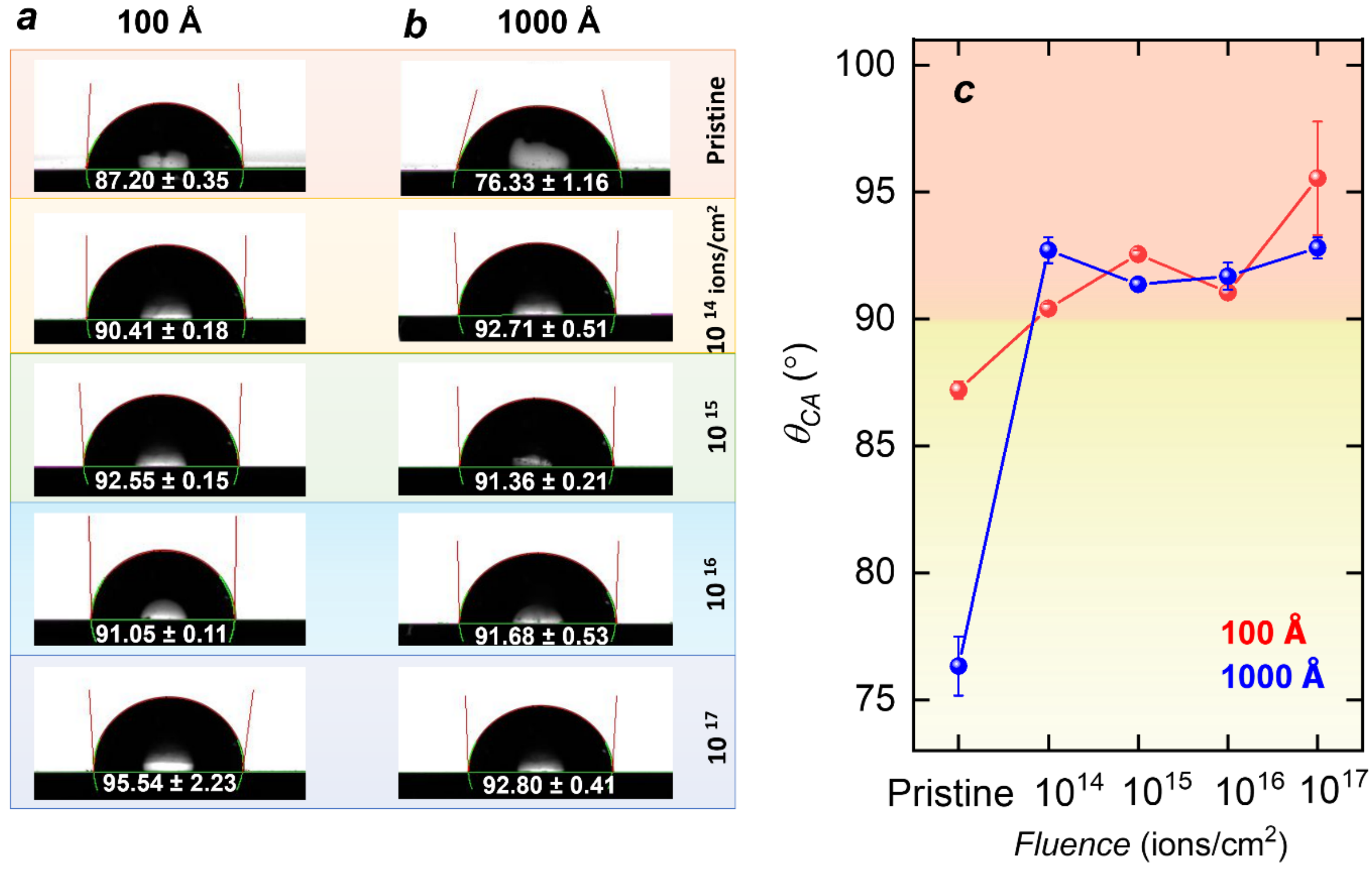

**Figure 6:** Drop shape analyser images depicting the wettability of Ru thin films with thicknesses of **(a)** 100 Å and **(b)** 1000 Å, in the pristine state and after $N^+$-ion irradiation at different fluences, where the values indicated in each image correspond to the measured contact angles. **(c)** Plot shows the contact angle ($\theta_{CA}$ in °)variation for Ru films having thicknesses of 100 Å and 1000 Å as a function on $N^+$-ion fluence.

## 4. Conclusion

In this study, the influence of $N^+$-ion irradiation on the structural and wettability properties of Ru thin films of varying thicknesses (100 Å and 1000 Å) was systematically examined over a wide fluence range ($10^{14}$-$10^{17}$ ions/cm$^2$). The findings clearly establish that ion irradiation is an effective tool for tailoring surface morphology, inducing significant modifications in both structural and functional properties. A consistent increase in surface roughness with ion fluence resulted in a corresponding enhancement in hydrophobicity, with the contact angle reaching a maximum of ~95.54° at the highest fluence. This transition from hydrophilic to hydrophobic behaviour is attributed to ion-induced surface texturing and defect formation, which alters the liquid-solid interactions. Overall, these findings highlight that controlled $N^+$-ion irradiation provides a tunable pathway for engineering thin films with desirable surface and structural characteristics. Such modifications are particularly beneficial for advanced electronic components and interconnect technologies, where enhanced hydrophobicity can reduce moisture adsorption and surface contamination, thereby improving device stability and performance.

**Acknowledgments**

The authors would like to thank CIF, LPU and CIC-UPES for providing the research infrastructure. The authors acknowledge the financial support from the UPES-SEED grant (UPES/R&D-SoAE/25062025/17).

**CRediT authorship contribution statement**

**Anmol Sharma:** Writing-review & editing, Writing- original draft, Formal analysis **Rajeev Kumar Gupta:** Writing-review & editing, Validation **Raj Kumar:** Resources, Writing-review & editing **Harsh Vardhan:** Resources, Writing-review & editing **Ratnesh Kumar Pandey:** Writing-review & editing, Formal analysis **Atul Thakur:** Resources, Writing-review & editing, **Shalendra Kumar:** Resources, Writing-review & editing **Ranjeet Kumar Brajpuriya:** Writing-review & editing, Visualization, Validation, Data curation **Vishakha Kaushik:** Writing-review & editing, Visualization, Validation, **Sachin Pathak:** Writing-review & editing, Writing-original draft, Visualization, Validation, Supervision, Methodology, Investigation, Formal analysis, Data curation, Conceptualization.

**Data availability**

Data will be made available on request.

**References:**

(1) Hu, S.; Lewis, N. S.; Ager, J. W.; Yang, J.; McKone, J. R.; Strandwitz, N. C. Thin-Film Materials for the Protection of Semiconducting Photoelectrodes in Solar-Fuel Generators. *J. Phys. Chem. C* **2015**, *119* (43), 24201–24228. https://doi.org/10.1021/acs.jpcc.5b05976.

(2) Xia, Y.; Halas, N. J. Shape-Controlled Synthesis and Surface Plasmonic Properties of Metallic Nanostructures. *MRS Bull.* **2005**, *30* (5), 338–348. https://doi.org/10.1557/mrs2005.96.

(3) Zhu, L.-Y.; Ou, L.-X.; Mao, L.-W.; Wu, X.-Y.; Liu, Y.-P.; Lu, H.-L. Advances in Noble Metal-Decorated Metal Oxide Nanomaterials for Chemiresistive Gas Sensors: Overview. *Nano-Micro Lett.* **2023**, *15* (1), 89. https://doi.org/10.1007/s40820-023-01047-z.

(4) Yu, E.; Kim, S.-C.; Lee, H. J.; Oh, K. H.; Moon, M.-W. Extreme Wettability of Nanostructured Glass Fabricated by Non-Lithographic, Anisotropic Etching. *Sci. Rep.* **2015**, *5* (1), 9362. https://doi.org/10.1038/srep09362.

(5) Hein, E.; Fox, D.; Fouckhardt, H. Self-Masking Controlled by Metallic Seed Layer during Glass Dry-Etching for Optically Scattering Surfaces. *J. Appl. Phys.* **2010**, *107* (3), 033301. https://doi.org/10.1063/1.3290969.

(6) Palumbo, F.; Lo Porto, C.; Favia, P. Plasma Nano-Texturing of Polymers for Wettability Control: Why, What and How. *Coatings* **2019**, *9* (10), 640. https://doi.org/10.3390/coatings9100640.

(7) Mukherjee, J.; Bhowmik, D.; Bhowmick, S.; Karmakar, P.; Bhattacharjee, S. Low Energy Ion-Beam Mediated Tailoring of Structural, Optical, and Electrical Properties of ITO Films. *Surf. Interfaces* **2025**, *59*, 105973. https://doi.org/10.1016/j.surfin.2025.105973.

(8) Brahma, R.; Ghanashyam Krishna, M. Ion Beam Sputtered Ultra-Thin and Nanostructured Ag Films for Surface Plasmon Applications. *Nucl. Instrum. Methods Phys. Res. Sect. B Beam Interact. Mater. At.* **2008**, *266* (8), 1493–1497. https://doi.org/10.1016/j.nimb.2008.02.017.

(9) Seo, J.; Jeong, H.; Kim, W.; Muñoz-García, J.; Castro, M.; Cuerno, R.; Kim, J.-S. Surface Nanopatterning of Si by Ion Beam Irradiation with Sub-Sputter-Threshold Energy. *Phys. Rev. B* **2025**, *112* (16), 165408. https://doi.org/10.1103/qkr5-2h54.

(10) Dhal, S.; Chatterjee, S.; Manju, U.; Tribedi, L. C.; Thulasiram, K. V.; Fernandez, W. A.; Chatterjee, S. Adhesive Hydrophobicity of $Cu_2$ O Nano-Columnar Arrays Induced by

Nitrogen Ion Irradiation. *Soft Matter* **2015**, *11* (47), 9211–9217. https://doi.org/10.1039/C5SM02142A.

(11) Aihaiti, L.; Tuokedaerhan, K.; Sadeh, B.; Zhang, M.; Xiang Qian, S.; Mijiti, A. Electrical and Microstructural Properties of Ta-C Thin Films for Metal Gate. *Mater. Res. Express* **2020**, *7* (7), 076402. https://doi.org/10.1088/2053-1591/aba0e9.

(12) Song, M.; Ameen, S.; Kim, D.-G.; Shin, H.-S.; Ansari, S. G.; Kim, Y.-S. Characterization of Ruthenium Thin Film on Tantalum by Electrochemical Deposition: Rutherford Backscattering Spectroscopy. *Sci. Adv. Mater.* **2011**, *3* (6), 932–938. https://doi.org/10.1166/sam.2011.1220.

(13) Sherif, E.-S. M. Corrosion Behavior of Duplex Stainless Steel Alloy Cathodically Modified with Minor Ruthenium Additions in Concentrated Sulfuric Acid Solutions. *Int. J. Electrochem. Sci.* **2011**, *6* (7), 2284–2298. https://doi.org/10.1016/S1452-3981(23)18184-3.

(14) Doná, E.; Cordin, M.; Deisl, C.; Bertel, E.; Franchini, C.; Zucca, R.; Redinger, J. Halogen-Induced Corrosion of Platinum. *J. Am. Chem. Soc.* **2009**, *131* (8), 2827–2829. https://doi.org/10.1021/ja809674n.

(15) Meng, X.; Deng, J.; Li, R.; Zhang, Q.; Tian, K.; Xu, J.; Yang, X.; Meng, L.; Du, J.; Wang, G. Effects of Ta Concentration on Microstructure, Optical and Optoelectronic Properties of Ga2O3:Ta Films. *Vacuum* **2024**, *224*, 113142. https://doi.org/10.1016/j.vacuum.2024.113142.

(16) Gao, K.; Zhang, Y.; Yi, J.; Dong, F.; Chen, P. Overview of Surface Modification Techniques for Titanium Alloys in Modern Material Science: A Comprehensive Analysis. *Coatings* **2024**, *14* (1), 148. https://doi.org/10.3390/coatings14010148.

(17) Zhao, M.; Chen, D.; Jin, J.; Kang, H.; Wang, Q.; Zhao, Z.; Zhou, Y.; Zhao, T. Hydrophobic Durability and Anti-Corrosion of Plasma Nitriding Layer: Evolution Mechanisms of Corrosion Behavior under Variations in Local Hydrochemistry. *Mater. Res. Express* **2023**, *10* (6), 066510. https://doi.org/10.1088/2053-1591/acde4a.

(18) Lee, K.; Hwang, W.; Cho, H. Development of a Versatile Coating Based on Hydrolysis-Assisted Self-Bonding and Structure Evolution of Aluminum Nitride Nanopowder: Application toward Repairing Severe Damages on Superhydrophobic Surfaces. *Surf. Coat. Technol.* **2023**, *460*, 129431. https://doi.org/10.1016/j.surfcoat.2023.129431.

(19) Du, X.; Gao, B.; Li, Y.; Song, Z. Super-Robust and Anti-Corrosive NiCrN Hydrophobic Coating Fabricated by Multi-Arc Ion Plating. *Appl. Surf. Sci.* **2020**, *511*, 145653. https://doi.org/10.1016/j.apsusc.2020.145653.

(20) Hu, Y.; Rasadujjaman, M.; Wang, Y.; Zhang, J.; Yan, J.; Baklanov, M. R. Study on the Electrical, Structural, Chemical and Optical Properties of PVD Ta(N) Films Deposited with Different N2 Flow Rates. *Coatings* **2021**, *11* (8), 937. https://doi.org/10.3390/coatings11080937.

(21) Priya, B.; Jasrotia, P.; Sulania, I.; Kumar, R.; Jyoti; Pandey, R. K.; Kumar, T. Substrate-Dependent Fractal Growth and Wettability of N+ Ion Implanted V2O5 Thin Films. *Appl. Surf. Sci.* **2023**, *619*, 156592. https://doi.org/10.1016/j.apsusc.2023.156592.

(22) Padhy, N.; Ningshen, S.; Panigrahi, B. K.; Kamachi Mudali, U. Corrosion Behaviour of Nitrogen Ion Implanted AISI Type 304L Stainless Steel in Nitric Acid Medium. *Corros. Sci.* **2010**, *52* (1), 104–112. https://doi.org/10.1016/j.corsci.2009.08.042.

(23) Wei, L.; Gao, Z. Recent Research Advances on Corrosion Mechanism and Protection, and Novel Coating Materials of Magnesium Alloys: A Review. *RSC Adv.* **2023**, *13* (12), 8427–8463. https://doi.org/10.1039/D2RA07829E.

(24) Serna-Manrique, M. D.; Escobar-Rincón, D.; Ospina-Arroyave, S.; Pineda-Hernández, D. A.; García-Gallego, Y. P.; Restrepo-Parra, E. Growth Mechanisms of TaN Thin Films Produced by DC Magnetron Sputtering on 304 Steel Substrates and Their Influence on the Corrosion Resistance. *Coatings* **2022**, *12* (7), 979. https://doi.org/10.3390/coatings12070979.

(25) Moyo, F.; Van Der Merwe, J. W.; Wamwangi, D. Corrosion Performance of Pulse Plated Ruthenium: Dependence on Pulse-off Time. *Surf. Coat. Technol.* **2016**, *307*, 971–977. https://doi.org/10.1016/j.surfcoat.2016.10.016.

(26) Moyo, F.; Van Der Merwe, J.; Wamwangi, D. Ruthenium Implantation for Corrosion Resistance: Using 2k Factorial Design to Determine Significant Implantation Parameters. *Results Surf. Interfaces* **2025**, *18*, 100465. https://doi.org/10.1016/j.rsurfi.2025.100465.

(27) Björck, M.; Andersson, G. *GenX* : An Extensible X-Ray Reflectivity Refinement Program Utilizing Differential Evolution. *J. Appl. Crystallogr.* **2007**, *40* (6), 1174–1178. https://doi.org/10.1107/S0021889807045086.

(28) Ziegler, J. F.; Ziegler, M. D.; Biersack, J. P. SRIM – The Stopping and Range of Ions in Matter (2010). *Nucl. Instrum. Methods Phys. Res. Sect. B Beam Interact. Mater. At.* **2010**, *268* (11–12), 1818–1823. https://doi.org/10.1016/j.nimb.2010.02.091.

(29) Mayr, S. G.; Averback, R. S. Effect of Ion Bombardment on Stress in Thin Metal Films. *Phys. Rev. B* **2003**, *68* (21), 214105. https://doi.org/10.1103/PhysRevB.68.214105.

(30) Baek, Min-Wook; 이용일. Synthesis and Catalytic Applications of Ruthenium(0) Nanoparticles in Click Chemistry. *Bull. Korean Chem. Soc.* **2014**, *35* (4), 1144–1148. https://doi.org/10.5012/BKCS.2014.35.4.1144.

(31) Gautam, N. K.; Alshoaibi, A.; Shaalan, N. M.; Dalela, S.; Alvi, P. A.; Sharma, A.; Brajpuriya, R. K.; Kumar, S. Novel Method of Synthesis for MXene/LIG Nanocomposites via Laser Fabrication Technique for Enhanced Performance in Energy Storage Applications. *J. Alloys Compd.* **2026**, *1057*, 186827. https://doi.org/10.1016/j.jallcom.2026.186827.

(32) Law, K.-Y. Definitions for Hydrophilicity, Hydrophobicity, and Superhydrophobicity: Getting the Basics Right. *J. Phys. Chem. Lett.* **2014**, *5* (4), 686–688. https://doi.org/10.1021/jz402762h.